\documentclass[10pt,journal,compsoc]{IEEEtran}

\usepackage{graphicx}
\usepackage{algorithm}
\usepackage[noend]{algpseudocode}
\usepackage{mathtools}
\usepackage{amsfonts}

\usepackage[bookmarks=true,breaklinks=true,colorlinks,linkcolor=black,citecolor=blue,urlcolor=black]{hyperref}

\begin{document}



\title{\huge A Survey of Machine Learning Applied to\\ Computer Architecture Design}


\author{\IEEEauthorblockN{
Drew D. Penney, and Lizhong Chen\IEEEauthorrefmark{1}, \textit{Senior Member, IEEE}\\
}
\thanks{\IEEEauthorrefmark{1}Corresponding author. Email: chenliz@oregonstate.edu}
\thanks{The authors are with Oregon State University, Corvallis, OR 97331}
\thanks{Copyright 2019 by Drew D. Penney and Lizhong Chen}
\thanks{All Rights Reserved}

}

\IEEEtitleabstractindextext{%
\begin{abstract}
Machine learning has enabled significant benefits in diverse fields, but, with a few exceptions, has had limited impact on computer architecture. Recent work, however, has explored broader applicability for design, optimization, and simulation. Notably, machine learning based strategies often surpass prior state-of-the-art analytical, heuristic, and human-expert approaches. This paper reviews machine learning applied system-wide to simulation and run-time optimization, and in many individual components, including memory systems, branch predictors, networks-on-chip, and GPUs. The paper further analyzes current practice to highlight useful design strategies and identify areas for future work, based on optimized implementation strategies, opportune extensions to existing work, and ambitious long term possibilities. Taken together, these strategies and techniques present a promising future for increasingly automated architectural design. 
\end{abstract}
}

\maketitle

\IEEEdisplaynontitleabstractindextext

\IEEEraisesectionheading{\section{Introduction}\label{sec:introduction}}

In the past decade, machine learning (ML) has rapidly become a revolutionary factor in many fields, ranging from commercial applications, as in self-driving cars, to medical applications, improving disease screening and diagnosis. In each of these applications, an ML model is trained to make predictions or decisions without explicit programming by discovering embedded patterns or relationships in the data. Notably, ML models can perform well in tasks/applications where relationships are too complex to model using analytical methods. These powerful learning capabilities continue to enable rapid developments in diverse fields. Concurrently, the exponential growth predicted by Moore's law has slowed, putting increasing burden on architects to supplant Moore's law with architectural advances. These opposing trends suggest opportunities for a paradigm shift in which computer architecture enables ML and, simultaneously, ML improves computer architecture, closing a positive-feedback loop with vast potential for both fields.

Traditionally, the relationship between computer architecture and ML has been relatively imbalanced, focusing on architectural optimizations to accelerate ML algorithms. In fact, the recent resurgence in AI research is, at least partly, attributed to improved processing capabilities. These improvements are enhanced by hardware optimizations exploiting available parallelism, data reuse, sparsity, etc. in existing ML algorithms. 
In contrast, there has been relatively limited work applying ML to improve architectural design, with branch prediction being one of a few mainstream examples. This nascent work, although limited, presents an auspicious approach for architectural design. 

\begin{figure}[t!]
	\centering
	\includegraphics[width=0.35\textwidth]{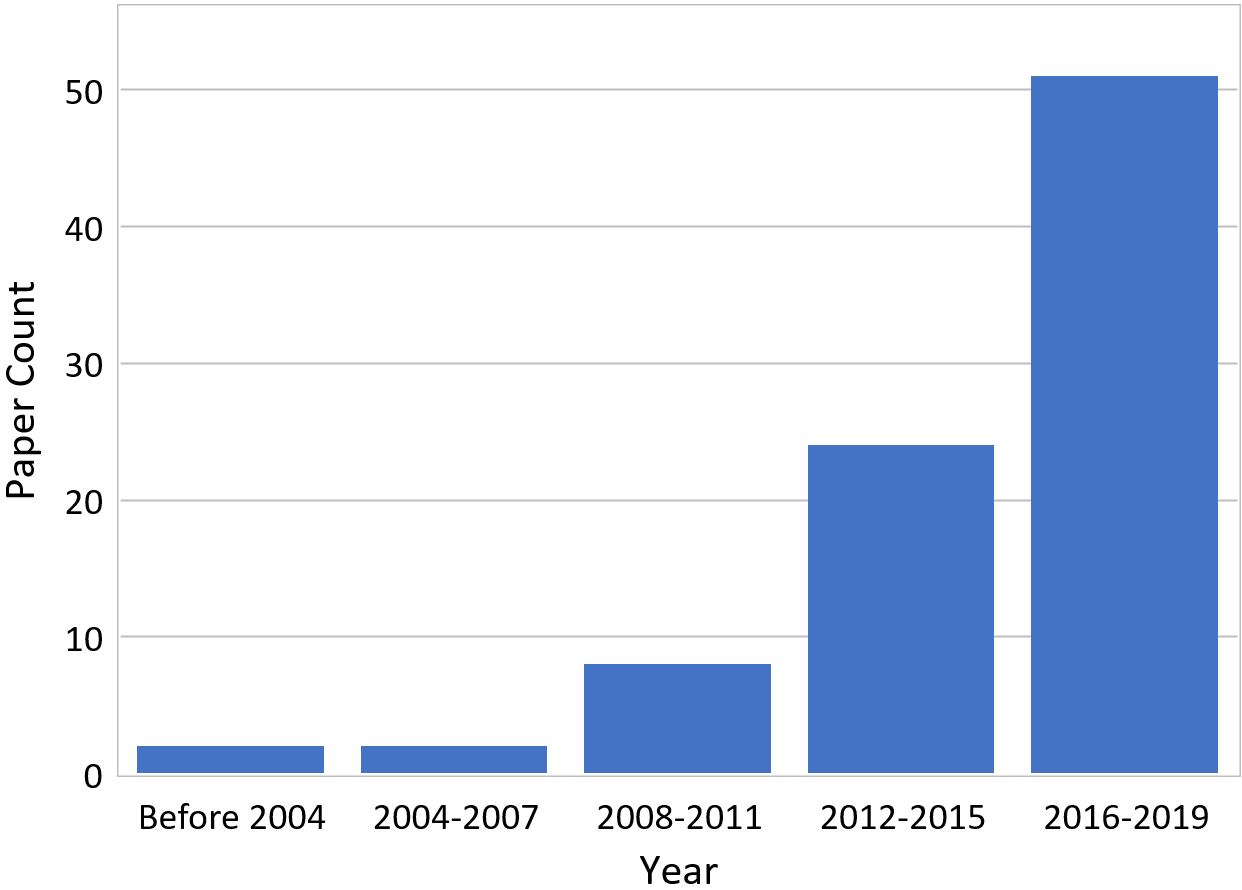}
	\caption{Publications on machine learning applied to architecture (for works examined in Section \ref{Overview})}
    \label{fig:PaperData}
    \vspace{-2.7ex}
\end{figure}

This paper presents an overview of ML applied to architectural design and analysis. As illustrated in Figure \ref{fig:PaperData}, this field has grown significantly in both success and popularity, particularly in the past few years. These works establish the broad applicability and future potential of ML-enabled architectural design; existing ML-based approaches, ranging from DVFS with simple classification trees to design space exploration via deep reinforcement learning, have already surpassed their respective state-of-the-art human expert and heuristic based designs. ML-based design will likely continue to provide breakthroughs as promising applications are explored.

The paper is organized as follows. Section 2 provides background on ML and existing models to build intuition on ML applicability to architectural issues. Section 3 presents existing work on ML applied to architecture. Section 4 then compares and contrasts implementation strategies in existing work to highlight significant design considerations. Section 5 identifies possible improvements and extensions to existing work as well as promising, new applications for future work. Finally, Section 6 concludes.

\section{Background}
\subsection{Fundamental Applicability}
Machine learning has been rapidly adopted in many fields as an alternative approach for a diverse range of problems. 
This fundamental applicability stems from the powerful relationship learning capabilities of ML algorithms. Specifically, ML models leverage a generic framework in which these models learn from examples, rather than explicit programming, enabling application in many tasks, including those too difficult to representing using standard programming methods. 
Furthermore, using this generic framework, there may be many possible approaches for any given problem. For example, in the case of predicting IPC for a processor, one can experiment with a simple linear regression model, which learns a linear relationship between features (such as core frequency and cache size) and the instructions-per-cycle (IPC). This approach may work well or it may work poorly. In the case it works poorly, one can try different features, non-linear feature combinations (such as core frequency times cache size), or a different model entirely, with another common choice being an artificial neural network (ANN). This diversity in possible approaches enables adjustment of models, model parameters, and training features to match the task at hand. 

\subsection{Learning Approaches \& Models}
The learning approach and the model are both fundamental considerations in applying machine learning to any problem. In general, there are four main categories of learning approaches: supervised learning, unsupervised learning, semi-supervised learning, and reinforcement learning. These approaches can be differentiated by \textit{what} data is used and \textit{how} that data is used to facilitate learning. Similarly, many appropriate models may exist for a given problem, thus enabling significant diversity in application based on the learning approach, hardware resources, available data, etc. In the following, we introduce these learning approaches and several significant models for each learning approach, focusing on approaches with proven applicability.
Implementation details are considered later in Section \ref{Analysis}. 

\textbf{Supervised learning}: In supervised learning, the model is trained using input features and output targets, with the result being a model that can predict the output for new, unseen inputs. Common supervised learning applications include regression (predicting a value such as processor IPC) and classification (predicting a label such as the optimal core configuration for application execution). Feature selection, discussed in Section \ref{sec:feat_sel}, is particularly important in these applications as the model must learn to predict solely based on feature values.

Supervised learning models can be generalized into four categories: decision trees, Bayesian networks, support vector machines (SVMs), and artificial neural networks \cite{EmergingML}. Decision trees use a tree structure where each node represents a feature and branches represent a value (or range of values) for that feature. Inputs are therefore classified by sequentially following branches based on the value of the feature being considered at a given node. Bayesian networks instead embed conditional relationships into a graphical structure; nodes represent random variables and edges represent conditional dependence between these variables. A performance prediction model, for example, can condition prediction for new applications on learned distributions for unobserved variables (i.e., underlying factors affecting performance) from other applications, as in \cite{LEO2015}. SVMs are generally known for their function rather than a particular graphical structure (as in decision tree and Bayesian networks). Specifically, SVMs learn the best dividing line (in 2-D) or hyperplane (in high dimensions) between examples, then uses examples along this hyperplane to make new predictions. SVMs can also be extended to non-linear problems using kernel methods \cite{StatLearning1999} as well as multi-class problems. 
Finally, artificial neural networks (or simply neural networks) represent a broad category of models that are, again, defined by their structure, which is reminiscent of neurons in the human brain; 
layers of nodes/neurons are connected using links with learned weights, enabling particular nodes to respond to specific input features. Simple perceptron models contain just one weight layer, directly converting the weighted sum of inputs into an output. More complex DNNs include several (or many) layers of these weighted sums. Additional variants such as convolutional neural networks (CNNs) incorporate convolution operations between some layers to capture spatial locality while recurrent neural networks re-use the previous output to learn sequences and long-term patterns. All these supervised learning models can be used in both classification and regression tasks, although there are some distinct high-level differences. Variants of SVMs and neural networks tend to perform better for high-dimension and continuous features and also when features may be nonlinear \cite{EmergingML}. These models, however, tend to require more data compared to Bayesian networks and decision trees.


\textbf{Unsupervised learning}: 
Unsupervised learning uses just input data to extract information without human effort. These models can therefore be useful, for example, in reducing data dimensionality by finding appropriate alternative representations or clustering data into classes that may not be obvious for humans. 

Thus far, the primary two unsupervised learning models applied to architecture are principal components analysis (PCA) and k-means clustering. PCA provides a method to extract significant information from a dataset by determining linear feature combinations with high variance \cite{PCA}. As such, PCA can be applied as an initial step towards building a model with reduced dimensionality, a highly desirable feature in most applications, albeit at the cost of interpretability (discussed in Section \ref{Analysis}). K-means clustering is instead used to identify groups of data with similar features. These groups may be further processed to generalize behavior or simplify representations for large datasets.

\textbf{Semi-supervised learning}: 
Semi-supervised learning represents a mix of supervised and unsupervised methods, with some paired input/output data, and some unpaired input data. Using this approach, learning can take advantage of limited labeled data and potentially significant unlabeled data. We note that this approach has, thus far, not yet found application in architecture. Nevertheless, one work on circuits analysis \cite{CoLearnIC2017} presents a possible strategy that could be adapted in future work. 

\textbf{Reinforcement Learning}: 
In reinforcement learning, an agent is sequentially provided with input based on an environment state and learns to perform actions that optimize a reward. For example, in the context of memory controllers, the agent replaces traditional control logic. Input could include pending reads and writes while actions could include standard memory controller commands (row read, write, pre-charge, etc.). Throughput could then be optimized by including it in the reward function. Given this setup, the agent will potentially, over time, learn to choose control actions that maximize throughput. 

Reinforcement learning models applied to architecture, as a whole, can be understood using a representation based on states, actions, and rewards. The agent attempts to learn a policy function $\pi$, which defines the action $a$ to take at a given state $s$, based on a received reward $r$ \cite{RLbook}. A learned state-value function, following the policy, is then given as
\begin{align}
\label{eq:optV}
  & V^\pi({s})= \mathbb{E} [\sum_{t\geq0}\gamma^{\: t}*r_t | s_0=s, \pi]
\end{align}
where $\gamma$ is a discount factor ($\leq 1$), which dictates how much the model should consider future rewards. The cumulative rewards are then maximized by learning an optimal policy $\pi^*$ that satisfies \begin{equation}
\label{eq:OptPolicy}
  \pi^*({s})= arg \max_\pi\mathbb{E} [\sum_{t\geq0}\gamma^t*r_t |  s_0=s, \pi].
\end{equation} 
Various models may implement different approaches to learn this optimal policy, but largely address the same problem of maximizing rewards. Q-learning is a noteworthy example that models an action-value function by estimating the value of an individual action, from a given state. 

\subsection{Feature Selection} \label{sec:feat_sel}
Supervised (and semi-supervised) learning methods rely upon input data features to model relationships and generate predictions. Consequently, approaches for feature selection can substantially impact model performance, including concerns such as over-fitting and computational overhead, as well as more abstract concerns, such as feature interpretability. In some works, feature selection is entirely based on expert knowledge. Additional, more general, approaches can either supplant or supplement expert knowledge.


One set of approaches, called filter methods, considers features individually using metrics involving statistical correlation or information theoretic criteria such as mutual information. These approaches are usually the least computationally intensive so may be preferred for very large feature sets, but model performance may be sub-optimal since evaluation criteria in filter methods do not consider feature context \cite{IntroFeatSel}; two features that provide little benefit individually may be beneficial together. Many alternative approaches therefore consider feature subsets. 

Wrapper methods provide a black-box method for feature selection by directly assessing the performance of a learning model \cite{IntroFeatSel}. Commonly applied greedy approaches include forward selection and backward elimination. In forward selection, features are progressively added to selected feature subset based on improvement to the overall learning model. Conversely, backward elimination removes features progressively that provide little benefit.

Embedded methods integrate feature selection into the learning model to provide a trade-off between filter and wrapper methods \cite{FeatSelSurvey}. Regularization is a widely used embedded method that allows the learning model to be fit, while simultaneously forcing feature coefficients to be small (or zero). Features with zero coefficient values can then be removed.  This method eliminates iterative feature selection present in wrapper methods, which can have high computational requirements \cite{IntroFeatSel}.


\section{Literature Review} \label{Overview}
This section reviews existing work that applies machine learning to architecture. Work is organized by sub-system (when applicable) or primary objective. We focus on design and optimization, but also introduce general performance prediction work.

\subsection{System Simulation}
Cycle-accurate simulators are commonly used in system performance prediction, but require several orders of magnitude more time than native execution. ML can offset this penalty through a trade-off between simulation time and accuracy. In general, ML can reduce execution time by 2-3 orders of magnitude with relatively high accuracy (task dependent, typically $>90\%$).
Early work by Ipek et al. \cite{ANNPred2006} modeled architectural design spaces using an ANN ensemble (a group of ANN predictors). Models were trained on approximately 1\% of the design space, then predicted CMP performance with 4-5\% error for random points, albeit only in that specific configuration space. When combined with SimPoints, predictions exhibit slightly higher error, but the simulated instruction count is further reduced.
Ozisikyilmaz et al. \cite{MLSysPredict2008} additionally predicted SPEC performance for future systems that may be poorly modeled by existing simulators. Evaluation was limited to randomly-sampled data with relatively simple linear regression and neural network models, but nevertheless demonstrated advantages for pruned neural networks compared to standard single-layer models (as in \cite{ANNPred2006}). 
Several other ML approaches have also been tested. Eyerman et al. \cite{MechEmpModel2011} proposed a mechanistic-empirical model for processor CPI prediction. In this approach, they used a generic mechanistic model with parameters inferred by a regression model. Their model is limited to single-core performance prediction, but improves accuracy, ease of implementation (compared to purely mechanistic models), and interpretability (compared to purely empirical models). 
Zheng et al. \cite{CrossPlatPred2015, PhasePred2016} explored cross-platform prediction from Intel/AMD to ARM processors using linear regression. Their first approach \cite{CrossPlatPred2015} made predictions based on a local neighborhood of examples around the target point to approximate non-linear behavior. They later \cite{PhasePred2016} emphasized phase-level prediction, assuming that phase-level behavior would be approximately linear. Notably, average error for cycle count predictions is less than 1\% using phase-level profiling. This approach is, however, restricted to a single target architecture and requires source code for phase-level analysis, leaving significant opportunities for future work.
Finally, recent work by Agarwal et al. \cite{ParallelPred2019} introduced a method to predict parallel execution speedup using single-threaded execution characteristics. They trained separate models for each thread count using application-level performance counters. Although neural networks were omitted due to limited data, evaluation found that Gaussian process regression still provided promising results, particularly for high thread counts.


\subsection{GPUs}

\textbf{Design Space Exploration}: GPU design space exploration has proven to be a particularly favorable application for ML due to a highly irregular design space; some kernels exhibit relatively linear scaling while others exhibit very complex relationships between configuration parameters, power, and performance \cite{Stargazer2012, GPUPerfPower2015, MultiObjGPU2016}. Jia et al. \cite{Stargazer2012} proposed Stargazer, a regression-based framework based on natural cubic splines. Stargazer randomly samples approximately 300 points from a target design space (933K points in evaluation) for each application, then applies stepwise regression on these points. Notably, the framework achieves under 3.8\% average performance prediction error.
Wu et al. \cite{GPUPerfPower2015} instead explicitly modeled scaling for compute units, core frequency, and memory frequency. Scaling data from training kernels was processed using k-means clustering to group kernels by scaling behavior. An ANN then classifies kernels into these clusters, allowing new kernels to be classified and predictions made using cluster scaling factors. This approach, in contrast to Jia et al. \cite{Stargazer2012}, therefore requires just a few samples for new applications. Jooya et al. \cite{MultiObjGPU2016}, similar to Jia et al. \cite{Stargazer2012}, considered a per-application performance/power prediction model, but additionally proposed a scheme to predict per-application Pareto fronts. Many ANN-based predictors were trained and the most accurate subset was used as an ensemble for prediction. Prediction accuracy was later improved by sampling points within a threshold of the previously predicted Pareto-optimal curve. Lin et al. \cite{MCPlace2019} combined a performance predicting DNN with a genetic search scheme to explore memory controller placement. The DNN was used as a surrogate fitness function, obviating slow system simulations. The resulting placement improves system performance by 19.3\%.

\textbf{Cross-Platform Prediction}: Porting applications for execution on GPUs is a challenging task with potentially uncertain benefits over CPU execution. Work has therefore examined methods to predict speedup or efficiency improvements using just CPU execution behavior. Baldini et al. \cite{GPUPred2014} cast the problem as a classification task, training a modified nearest-neighbor and a support vector machine (SVM) model to determine, based on a threshold, whether GPU implementation would be beneficial. Using this approach, they predicted near-optimal configurations 91\% of the time. In contrast, Ardalani et al. \cite{XAPP2015} trained a large ensemble of regression models to directly predict GPU performance for the code segment. Although several code segments exhibit high error, the geometric mean of the absolute value of the relative error is still just 11.6\% and the model successfully identifies several code segments (both beneficial and non-beneficial) that are incorrectly predicted by human experts. Later work by Ardalani et al. \cite{SXAPP2019} introduced a completely static-analysis-based framework using a random forest model for binary classification. This approach eliminates both dynamic profiling and human guidance, instead using features such as instruction mix, branch divergence estimates, and kernel size to provide 94\% accuracy for binary speedup classification (using a speedup threshold of 3). 

\textbf{GPU Specific Prediction \& Classification}: O'Neal et al. \cite{HALWPE2017} presented a methodology for next-generation GPU performance prediction as cycles-per-frame (CPF) for DirectX applications. They focused on Intel GPUs, profiling earlier-generation architectures (e.g., Haswell GT2) to train next-generation predictors. They found that different models (i.e., linear vs non-linear) can produce more accurate results depending on the prediction target (Broadwell GT2/GT3 vs Skylake GT3), with the best performing models achieving less than 10\% CPF prediction error. Recent work by Li et al. \cite{GPUTrafficPattern2019} presented a re-evaluation of commonly accepted knowledge of GPU traffic patterns. They used a CNN and t-distributed stochastic neighbor embedding on heatmap-transformed traffic data, identifying eight unique patterns with 94\% accuracy. 

\textbf{Scheduling}: 
GPU processing-in-memory (PIM) architectures can benefit from high memory bandwidth with reduced data movement energy. Despite this benefit, potential limitations on PIM compute capabilities may introduce complex trade-offs between performance and energy when scheduling execution on various resources.
For this reason, Pattnaik et al. \cite{GPUPIM2016} proposed an approach using a regression model to classify core affinity, thus dividing the workload, and an additional regression model to predict execution time, enabling dynamic task migration. Performance and energy efficiency are improved by 42\% and 27\%, respectively, over a baseline GPU architecture. Further improvements are possible by improving core affinity classification accuracy (compared to regression).

\subsection{Memory Systems and Branch Prediction}
\textbf{Caches}: Heuristic approaches for caching can incur performance penalties due to dramatic workload variance. ML approaches can learn these intricacies and offer superior performance. Peled et al. \cite{Semantic2015} proposed a prefetcher exploiting semantic locality (data structures) using contextual bandits (a simple RL variant), correlating contextual information and candidate addresses for prefetching. Implementation uses a two-level indexing method to dynamically control state information, allowing online feature selection with some additional overhead. Zeng and Guo \cite{LSTMPrefetch2017} proposed a long short-term memory (LSTM) model (a recurrent neural network variant)
for prefetching based on local history and offset-delta tables. Evaluation showed that the LSTM model enables accurate predictions over longer sequence and higher noise resistance than prior work. Several concerns relating to overhead and warm-up time are addressed, with potential solutions remaining for future work. Similarly, Braun et al. \cite{LSTMPrefetchPatterns2019} extensively explored LSTM prefetching accuracy under several common access patterns. Experiments considered the impact of lookback size (access history window) and LSTM model size for several noise levels and independent access stream counts. Recent work by Bhatia et al. \cite{PPF2019} synthesized traditional prefetchers with a perceptron-based prefetch filter, allowing aggressive predictions without degrading accuracy. Evaluation confirmed substantial coverage and IPC benefits offered by the proposed scheme, with 9.7\% IPC speedup over the next best prefetcher when referenced to a no-prefetching four-core baseline. ML has similarly been applied to data reuse policies. For example, Teran et al. \cite{Perceptron2016} predicted LLC reuse with a perceptron model. In this approach, input features are hashed to access saturating weight tables that are incremented/decremented based on correct/incorrect reuse prediction. These features are chosen empirically and shown to significantly impact performance, thus presenting an option for further optimization. Wang et al. \cite{SSDCacheML2018} predicted reuse prior to cache entry, only storing data in the cache if there was predicted reuse. They used decision trees as a low-cost alternative to ensemble models, achieving 60-80\% reduction in writes. Additional research has explored the growing performance bottleneck in translation lookaside buffers (TLBs). Margaritov et al. \cite{TLBPred2018} proposed a scheme for virtual address translation in TLBs based on learned indices \cite{LearnedIndexes2018}. Evaluation showed nearly 100\% accuracy for predicted indices, but practical implementation will require dedicated hardware to reduce calculation overhead (and is left for future work). 

\textbf{Schedulers \& Control}: Controllers for memory and storage systems influence both device performance and reliability, thus representing another strong application for ML models compared with heuristics. Ipek et al. \cite{SelfOptMC2008} first proposed an RL approach for memory controllers to capture the balance between concurrency, delay, and several other factors. The model predicted optimal actions (precharge, activate, row read/write), improving system performance by 15\% (in a two-channel system) over prior work. Mukundan and Martinez \cite{MORSE2012} later built upon Ipek's work, generalizing the reward function to optimize energy, fairness, etc. They also added power-up and power-down actions to enable a further 8.6\% improvement in performance and a significant improvement in energy efficiency. Related work optimizes communication energy between memory/storage and other systems using ML. Manoj et al. \cite{TSIVoltage2016} proposed a Q-learning method for dynamic voltage swing control in through-silicon-interposer transmission lines. Predictions for power and bit error rate were quantized, then provided as input to the model to determine a new voltage level. Although their approach requires significant quantization to minimize overhead, they still achieved 15.1\% energy savings compared to a static voltage baseline. Wang and Ipek \cite{DataCluster2016} reduce data movement energy through online clustering and encoding. Several clusters are continuously updated at a bit-level using majority voting for data in that cluster. The total number of transmitted 1s is then minimized by XORing new data with the closest learned cluster center. Kang and Yoo \cite{RLTailLatency2018} applied Q-learning to manage garbage collection in SSDs by determining optimal periods of inactivity. Key states are kept in the Q-table using LRU replacement, allowing a vast state space and, ultimately, a 22\% average tail latency reduction over the baseline. Many states are, however, observed only once per workload, suggesting potential benefits using deep Q-learning (DQL). Other work directly considered system reliability. For example, Deng et al. \cite{MCT2017} proposed a regression-based framework to dynamically optimize performance and lifetime in non-volatile memories. Their approach used phase-based application statistics to manage several conflicting policies for write latency, write cancellation, endurance, etc., guaranteeing a minimum lifetime with modest performance/energy improvements. Xiao et al. \cite{DiskFailPred2018} proposed a method for disk failure prediction using an online random forest. They trained their model using a disk status window to account for imprecision in recorded failure date, enabling accurate predictions of soon-to-be faulty drives. Comparison against other random forest updating schemes (e.g., updating once a month) highlighted accuracy benefits from consistent training that may be extended to related domains. 

\textbf{Branch Prediction}: Branch prediction is a noteworthy example of current ML application in industry, with accuracy surpassing prior state-of-the-art non-ML predictors. The perceptron-based branch predictor was first proposed by Jim\'enez and Lin \cite{PerceptronPred2001} as a promising high-accuracy alternative to two-level schemes using pattern history tables. Later research by St. Amant et al. introduced SNAP \cite{AnalogNeuralPred2008}, a perceptron-based predictor implemented using analog circuitry to enable an efficient and practically feasible design. Perceptron weights and branch history were used to drive current-steering DACs that perform the dot product as the sum of currents. Jim\'enez \cite{OptNeuralPred2011} further optimized this design using a per-branch history table, dynamic coefficients for history importance, and a dynamic learning threshold. The optimized design achieves 3.1\% lower MKPI than L-TAGE. Recent work with perceptron-based predictors by Garza et al. \cite{BLBP2019} explored bit-level prediction for indirect branches. Possible branch targets are evaluated using their similarity (dot product) with the combined weights from eight feature tables incorporating local and global history, ultimately reducing MKPI by 5\% compared to ITTAGE. Currently, state-of-the-art conditional branch predictors (e.g., TAGE-SC-L \cite{TAGESCL}) still hide significant IPC gains (14.0\% for an Intel Skylake architecture) in just a few hard-to-predict (H2P) conditional branches \cite{CNNHelperPred2019}. Tarsa et al. \cite{CNNHelperPred2019} consequently proposed ``CNN Helper'' predictors that target specific H2Ps using simple two-layer CNNs. Results indicate strong applicability across diverse workloads and present a promising area for future work.

\subsection{Networks-on-Chip}
\textbf{DVFS \& Link Control}: Modern computing systems exploit complex power control schemes to enable increasingly parallel architectural designs. Heuristic schemes may fail to exploit all energy-saving opportunities, particularly in dynamic network-on-chip (NoC) workloads, leading to significant benefits through proactive ML-based control. Savva et al. \cite{ANNNoCLinks2012} implemented dynamic link control using several ANNs, each of which monitors a NoC partition. These ANNs used just link utilization to learn a dynamic threshold to enable/disable links. Despite energy savings, their approach can cause high latency under dimension-ordered routing. DiTomaso et al. \cite{LESSON2017} relocated flit buffers to the links and dynamically controlled both link direction and power-gating with per-router classification trees. Using a simple three-level tree to limit overhead, overall NoC power is reduced by 85\% and latency is reduced by 14\% compared to a concentrated mesh. Winkle et al. \cite{Photonic2018} explored ML-based power scaling in photonic interconnects. Even a simple linear regression model provided promising results, negligibly reducing throughput (versus no power-gating) while reducing laser power consumption by 42\%. Reza et al. \cite{NeuroNoC2018} proposed a multi-level ANN control scheme that considered both power and thermal constraints on task allocation, link allocation, and node DVFS. Individual ANNs classified appropriate configurations for local NoC partitions while a global ANN classified optimal overall resource allocation. This scheme identifies the global optimal NoC configuration with high accuracy (88\%), but uses complex ANNs that could impact implementation. Clark et al. \cite{LEAD2018} proposed a router design for DVFS and evaluated several regression-based control strategies. Variants predicted buffer utilization, change in buffer utilization, or a combined energy and throughput metric. This work was expanded by Fettes et al. \cite{LEADRL2019}, who introduced an RL control strategy. Both regression and RL models enable beneficial tradeoffs, although the RL strategy is most flexible.

\textbf{Admission \& Flow Control}: As with NoC DVFS, both admission and flow control can benefit from proactive prediction. Early work by Boyan and Littman \cite{Boyan94} introduced Q-learning based routing in networks using delivery time estimates from neighboring nodes, noting throughput advantages over traditional shortest path routing for high traffic intensity. Several works have expanded upon Q-routing, observing application in dynamically changing NoC topologies \cite{Majer2005}, improved capabilities in bufferless NoC fault-tolerant routing \cite{Feng2010}, and high-performance congestion-aware non-minimal routing \cite{Ebrahimi2012}. More recent works have instead focused on injection throttling and hotspot prevention. 
For example, Daya et al. \cite{SCEPTRE2016} proposed SCEPTER, a bufferless NoC using single-cycle multi-hop paths. They controlled injection throttling using Q-learning to maximize multi-hop performance and improve fairness by reducing contending flits. Future work could reduce Q-table overhead which scales with NoC size in their implementation. 
Wang et al. \cite{ANNAdmission2019} instead used an ANN to predict optimal injection rates for a standard buffered NoC. Additional preprocessing (to capture both spatial and temporal trends) and node grouping enables high accuracy predictions (90.2\%) and reduces execution time by 17.8\% compared to an unoptimized baseline. Soteriou et al. \cite{ANNHotspot2015} similarly explored ANN-based injection throttling to reduce NoC hotspots. The ANN was trained to predict hotspots while recognizing the impact of proposed injection throttling and dynamic routing, providing a holistic mitigation strategy. 
The model provides state-of-the-art results for throughput and latency under synthetic traffic, but limited improvement under real-world benchmarks, suggesting the potential for further optimization. 
Another Q-learning approach, proposed by Yin et al. \cite{DRLArbitration2018}, used DQL to arbitrate NoC traffic. They considered a wide range of features and rewards while noting that the proposed DQL algorithm is impractical due to overhead. Regardless, evaluation exhibited modest throughput improvements over round-robin arbitration. 

\textbf{Topology \& General Design}: Several works also applied ML to higher-level NoC topology design, involving trade-offs between power and performance, with some further considering thermals. Das et al. \cite{3DNoC2015} used a ML-based $STAGE$ algorithm to efficiently explore small-world inspired 3D NoC designs. In this approach, design alternates between base/local search (adding/removing links in a hill-climbing approach) and meta search (predicting beneficial starting points for local search using prior results). The same model was used again by Das et al. \cite{Energy3DNoC2016} to balance link utilization and address TSV reliability concerns. The STAGE algorithm was then enhanced by Joardar et al. \cite{MOOSTAGE2018} to optimize a heterogeneous 3D NoC design. The models explores multi-objective trade-offs between CPU latency, GPU throughput, and thermal/energy constraints. All three works \cite{3DNoC2015, Energy3DNoC2016, MOOSTAGE2018} still rely upon hill-climbing for optimization. Recent work by Lin et al. \cite{DRL2019} instead explored deep reinforcement learning in routerless NoC design. They used a Monte Carlo tree search to efficiently explore the search space and a deep convolutional neural network to approximate both the action and policy functions, thereby optimizing loop configurations. Further, the proposed deep reinforcement learning framework can strictly enforce design constraints that may be violated by prior heuristic or evolutionary approaches.
Rao et al. \cite{MLNoC2018} investigated multi-objective NoC design optimization across a broad SoC feature space (from bandwidth requirements to SoC area). ML models were trained using data from thousands of SoC configurations to predict optimal NoC designs based on performance, area, or both. Limited comparisons against human-expert designs did not consider alternative techniques (e.g., AMOSA \cite{AMOSA2008}), yet exhibited some promising results, motivating research into effective features and models as well as further comparisons against alternative techniques.

\textbf{Performance Prediction}: Existing NoC models based on queuing theory are generally accurate, but rely on assumptions of traffic distribution that may not hold for real applications \cite{SVR-NoC2013}. Qian et al. \cite{SVR-NoC2013} emphasized how ML-based approaches can relax the assumptions made by queueing theory models. They constructed a mechanistic-empirical model based on a communication graph, using support vector regression (SVR) to relate several features and queuing delays. Evaluation showed lower error (3\% error vs 10\% error) than an existing analytical approach. 
Sangaiah et al. \cite{UncoreRPD2016} considered both NoC and memory configuration for performance prediction and design space exploration. Following a standard approach, they sampled a small portion of the design space, then trained a regression model to predict the resulting system CPI. Evaluation generally showed high accuracy, but lower accuracy for high-traffic workloads (median error of 24\%). Additional design space exploration exhibited promising results, reducing the design space from 2.4M points to less than 1000.

\textbf{Reliability \& Error Correction}: Overhead introduced by error correction in NoCs can be significant, especially when re-transmission is required. Several works have, therefore, explored ML-based control schemes. DiTomaso et al. \cite{ErrorNoC2016} trained a decision tree to predict NoC faults using a wide range of parameters including temperature, utilization, and device wear-out. These predictions allow proactive encoding (on top of the baseline cyclic redundancy check) for transmission that are likely to have errors. Wang et al. \cite{FaultRLNoC2019} adopted a similar strategy for dynamic error mitigation, but used an RL-based control policy to eliminate the need for labeled training examples. Their approach provides an average of 46\% dynamic power savings (17\% better than the decision tree method \cite{ErrorNoC2016}) compared with a static CRC scheme. In both cases, ML-based proactive control chose a  more efficient scheme than CRC only. Wang et al. \cite{IntelliNoC2019} subsequently proposed a holistic framework for NoC design incorporating dynamic error mitigation, router power-gating, and multi-function adaptive channel buffers (MFAC buffers). They emphasized comprehensive benefits through synergistic integration/control of several architectural innovations, thus achieving substantial improvements in latency (32\%), energy-efficiency (67\%), and reliability (77\% higher Mean Time to Failure) compared to a SECDED baseline.

\subsection{System-level Optimization}

\textbf{Energy Efficiency Optimization}: Significant work has begun to consider systems in which workload execution is constrained by total energy consumption rather than processing resources. Control schemes incorporating ML have shown promise in optimizing energy efficiency with minimal performance reduction, often enabling 60-80\% reductions in the energy-delay product compared to race-to-idle schemes. Won et al. \cite{BootstrapCMP2014} introduced a hybrid ANN + PI (proportional-integral) controller scheme for uncore DVFS. They initially trained the ANN offline, then refined predictions online using the PI controller. This hybrid scheme was shown to reduce the energy-delay product by 27\% compared to a PI controller alone, with less than 3\% performance degradation compared to the highest V/F level. Pan et al. \cite{MLRL2014} implemented a power management scheme using a multi-level RL algorithm. Their method propagates individual core states up a tree structure while aggregating Q-learning representations at each level. Global allocation is made at the root, then decisions are propagated back down the tree, enabling efficient per-core control. Bailey et al. \cite{AdaptiveHeterogeneous2014} addressed power efficiency in heterogenous systems. Similar to Wu et al. \cite{GPUPerfPower2015}, they clustered kernels by their scaling behavior to train multiple linear regression models. Runtime prediction used two sample configurations, one from CPU execution and one from GPU execution, to determine the optimal configuration. Lo et al. \cite{RuntimeDVFS2015} focused on energy-efficiency optimization for real-time interactive workloads. They used linear regression to model execution time based on annotations and code features, enabling stricter service level guarantees at the cost of applicability when source code is unavailable. Mishra et al. \cite{CALOREE2018} also addressed real-time workloads, combining control theory and several ML-based models. Their framework was realized by offloading learning to a server, allowing low overhead DVFS that reduces energy consumption by 13\% compared to the best prior approach. Related work by Mishra et al. \cite{LEO2015} applied a comparatively complex hierarchical Bayesian model to combine both offline and online learning. In this approach, they accepted a high execution time penalty (0.8s) in order to provide significantly more accurate predictions than online or offline training alone. This approach therefore targeted longer executing workloads, but can provide more than 24\% energy savings over the next best approach. Bai et al. \cite{VoltRegRL2017} implemented a RL-based DVFS control policy adapted to a novel voltage regulator hierarchy using off-chip switching regulators and on-chip linear regulators. Individual RL agents adapt to a dynamically allocated power budget determined by a heuristic bidding approach. The design was enhanced using adaptive Kanerva coding \cite{KanervaCode} to limit area/power overhead and experience sharing to accelerate learning. Chen and Marculescu \cite{DistributedRL2015} (later Chen et al. \cite{PROFIT2018}) explored an alternative two-level strategy for RL-based DVFS. Similar to Bai et al. \cite{VoltRegRL2017}, they used RL agents at a fine-grain core level to select a V/F level based on an allocated share of the global power budget. They achieved further improvement by allocating power budget using a performance-aware, albeit still heuristic-based, variant that considers relative application performance requirements. 
Imes et al. \cite{MLScheduling2018} explored single-application system energy optimization for a broader range of configurations options including socket allocation, HyperThread usage, and processor DVFS. They identified several useful models, while noting that further work could optimize models and parameters. Analysis also provided insight into the benefit from single-model multi-resource optimization, particularly for neural networks. Finally, recent work by Tarsa et al. \cite{CPUAdapt2019} considered an ML framework for post-silicon CPU adaptations using firmware updates to microcontroller-implemented models. Significant accommodations for statistical blindspots limit the rate of service-level-agreement violations while optimizing performance per watt for both general-purpose and application-specific deployment.

\textbf{Task Allocation and Resource Management}: In addition to energy control, ML offers an approach to allocate resources to tasks or tasks to resources by predicting the impact of various configurations on long-term performance. 
Lu et al. \cite{RLThermalAlloc2015} proposed a thermal-aware Q-learning method for many-core task allocation. The agent considered only current temperature (i.e., no application profiling or hardware counters), receiving higher rewards for task assignments resulting in greater thermal headroom. Evaluation indicated an average 4.3$^{\circ}$C reduction in peak temperature compared to a heuristic approach.
Nemirovsky et al. \cite{HeterogeneousML2017} introduced a method for IPC prediction and task scheduling on a heterogeneous architecture. They predicted IPC for all task arrangements using ANNs, then selected the arrangement with the highest IPC. Evaluation highlighted significant throughput gains ($>1.3x$) using a deep (but high overhead) neural network, indicating one possible application for pruning (discussed in Section \ref{sec:strats}).
Recent work has also explored multi-level scheduling in hybrid CPU-GPU clusters. Zhang et al. \cite{VideoDRL2018} proposed a deep reinforcement learning (DRL) framework to divide video workloads, first at the cluster level (selecting a worker node) and then at the node level (CPU vs GPU). The two DRL models act separately, but still work together to optimize overall throughput. 
Allocating resources to tasks is another possible approach. Early work by Bitirgen et al. \cite{ManageCMP2008} considered a system with four cores and four concurrent applications. In their approach, per-application ANN ensembles predicted IPC for 2,000 configurations at each interval (500K cycles). IPC predictions were then aggregated to choose the highest performing overall system configuration. Scaling concerns for per-application ensembles and exponentially increasing configuration spaces could be addressed in future work.
Recent research has also considered low-level co-optimization involving multiple components/resources. For example, Jain et al. \cite{MLM2016} explored concurrent optimization of core DVFS, uncore DVFS, and dynamic LLC partitioning. These options are optimized by individual agents (potentially limiting co-optimization opportunities) at a relatively large interval (1B instructions). Evaluation nevertheless indicated noteworthy reductions in energy-delay-product through multi-resource optimization.
Finally, work by Ding et al. \cite{GenMPLearn2019} established a somewhat contradictory trend between model accuracy and system optimization goals based on improvements for data scarcity and model bias. Specifically, they found that state-of-the-art models exhibit diminishing returns for accuracy and instead benefit from domain knowledge (e.g., focus sampling on the optimal front). 

\textbf{Chip Layout}: Work by Wu et al. \cite{FFCluster2016} demonstrated uses for ML in chip layout, deviating from the common applications including control, prediction, and design space exploration. They used k-means to cluster flip-flops during physical layout, minimizing clock wirelength at the expense of signal wirelength, noting that clock networks can consume more than 40\% of chip power. They included constraints on maximum flip-flop displacement and cluster size, generating designs with 28.3\% reduced displacement, 3.2\% reduced total wirelength, and 4.8\% reduced total switching power compared to the prior state-of-the-art approach.

\textbf{Security}: Malware detection, a traditionally software-based task, has been explored using machine learning to enable reliable hardware-based in-execution detection. For example, Ozsoy et al. \cite{MalwareDetect2016} test both logistic regression (LR) and neural network classifiers trained on low-level hardware counters. Optimization based on reduced precision and feature selection provides high accuracy (100\% malware detection and less than 16\% false positives) with minimal overhead (0.04\% core power and 0.19\% core logic area) for the LR model.

\subsection{ML-Enabled Approximate Computing} \label{MLapprox}
Approximate computing has many facets, including circuit level approximation (such as reduced precision adders), control level approximation (relaxing timings, etc), and data level approximation. Methods using ML generally fall within this last category, offering a powerful function/loop approximation technique that commonly provides 2-3 times application speedup and energy reduction with limited impact on output quality. Esmaeilzadeh et al. \cite{NPU2012} introduced NPU, a new approach to programmable approximation using neural networks. They developed a framework to realize Parrot transformations that translate annotated code segments into neural networks approximations. Tightly integrating the NPU with the CPU allowed an average 2.3x speedup and 3.0x energy reduction in studied applications. This framework was later extended by Yazdanbakhsh et al. \cite{NGPU2015} to implement neural approximation on GPUs. Neural approximation was integrated into the existing GPU pipeline, enabling component re-use and approximately 2.5x speedup and 2.5x reduced energy. Grigorian et al. \cite{BRAINIAC2015} presented a different approach for a multi-stage neural accelerator. Inputs are first sent through a relatively low accuracy/overhead neural accelerator, then checked for quality; acceptable results are committed, while low quality approximations are forwarded to an additional, more precise, approximation stage. The problem with these works is that error is either constant \cite{NPU2012, NGPU2015} or requires several stages with potentially redundant approximation \cite{BRAINIAC2015}. For that reason, Mahajan et al. \cite{MITHRA2016} introduced MITHRA, a co-designed hardware-software control framework for neural approximation. MITHRA implements configurable output quality loss with statistical guarantees. ML classifiers predict individual approximation error, allowing comparison to a quality threshold. Recent work by Oliveira et al. \cite{ApproxClassification2018} also explored approximation using low-overhead classification trees. Even with software-based execution, they achieved application speedup comparable to an NPU \cite{NPU2012} hardware implementation. Finally, ML has also been used to mitigate the impact of faults in existing approximate accelerators. Taher et al. \cite{MLFaultApprox2018} observed that faults tend to manifest in a similar manner across many input test vectors.
This observations enables effective error compensation using a classification/regression model to correct output based on predicted faults for a given input.

\section{Analysis of Current Practice}\label{Analysis}

This section examines varying techniques employed in existing work. These comparisons emphasize potentially useful design practices and strategies for future work.

Work is divided into two categories that represent a natural division in design constraints and operating timescales and therefore correspond to differing design practices. The first category, online ML application, encompasses work that directly applies ML techniques at run-time, even if training is performed offline. Design complexity in this work is therefore inherently limited by practical constraints such as power, area, and real-time processing overhead. The second category, offline ML application, instead applies ML to guide architectural implementation, involving tasks such as design and simulation. Consequently, models for offline ML application can exploit higher complexity and higher overhead options at the cost of training/prediction time.

\subsection{Online ML Application} \label{OnlineApp}

\textbf{Model Selection}: 
Online ML applications primarily use either decision trees or ANNs, in the case of supervised learning models, and either Q-learning or deep Q-learning, in the case of RL models. Note that tasks for these learning approaches are not necessarily disjoint, particularly for control. Fettes et al. \cite{LEADRL2019} cast DVFS as both a supervised learning regression task and as a reinforcement learning task. The supervised learning approach predicted buffer utilization or change in buffer utilization to determine an appropriate DVFS mode. In contrast, the RL approach directly used DVFS modes as the action space. Both models can perform well, but the RL model is more universally applicable since the energy/throughput trade-off can be tailored to application needs and does not require threshold tuning. This certainly does not mean that RL is a one-model-fits-all solution. Supervised learning models find strong application in function approximation \cite{NPU2012, NGPU2015, BRAINIAC2015, ApproxClassification2018} and branch prediction tasks \cite{AnalogNeuralPred2008, OptNeuralPred2011}, which are far less suitable (if not impossible) to approach using RL since these tasks cannot be represented well as a sequence of actions.

\textbf{Implementation \& Overhead}:
Implementation of online ML applications highlight limitations in data availability, storage space for models, etc., indicating the need for an efficient, and generally low complexity, model. These limitations will likely become more important to consider as more research moves towards real-world implementation.

Several NoC-based works \cite{ANNNoCLinks2012, SCEPTRE2016, BootstrapCMP2014} 
have applied different methods for global data collection to support ML models. 
Daya et al. \cite{SCEPTRE2016} implemented self-learning injection throttling using a separate bufferless starvation network that carries a starvation flag, encoded as a one-hot N-bit vector for a network with N nodes. These starvation vectors are propagated to all nodes, allowing individual node-based Q-learning agents to determine appropriate injection throttling. Soteriou et al. \cite{ANNHotspot2015} similarly used a dedicated networks to collect buffer utilization and VC occupancy statistics. The ANN-based DVFS control proposed by Won et al. \cite{BootstrapCMP2014} eschewed an additional status/data network by encoding data into unused bits in standard packet headers. Data is opportunistically collected by a central control unit as packets pass through its router. This method introduces potential concerns about data staleness, but prior work \cite{NoCMonitor2012} observed nearly identical performance to omniscient data collection for sufficiently large (50K cycle) control windows. Smaller time windows can be accommodated by sending dedicated packets, as done by Savva et al. \cite{ANNNoCLinks2012}. 


Implementation can also consider the use of either hardware or software models. Implementation using dedicated hardware will usually experience lower execution time overhead, but there are other considerations. Esmaeilzadeh et al. \cite{NPU2012} implemented a neural processor (NPU) for function approximation using a dedicated hardware module. They also considered a software implementation, but observed a prohibitive increase in instruction count for software execution compared to a baseline x86 function. Later work by Oliveira et al. \cite{ApproxClassification2018} found that function approximation using a simple classification tree can achieve comparable results to NPU \cite{NPU2012} for application speedup and error rate  in several applications (albeit somewhat worse on average). Their purely software implementation highlights a trade-off between area/power and accuracy/performance. Won et al. \cite{BootstrapCMP2014} observed a similar trade-off, choosing to implement an ANN in software using an on-die microcontroller rather than dedicated hardware. This implementation consumes several orders of magnitude more cycles (15K cycles for inference), but requires 50mW less average power than a hardware implementation.

Approaches for hardware implementation may also vary based on the task. A ``standard'' ANN implementation is observed in work by Savva et al. \cite{ANNNoCLinks2012}. They incorporated a finite state machine for control, an array of multiply-accumulate (MAC) units for calculation, a register array to load and store results, and a lookup-table-based activation function. Both MAC array width and calculation precision can be adjusted to balance power/area and accuracy/speed. In contrast, St. Amant et al. \cite{AnalogNeuralPred2008} implemented a perceptron branch predictor using a mixed signal design. They realized dot products in analog circuitry, leveraging transistor sizing and current summing to achieve a feasible overhead. Variance also exists in hardware for RL models. The ``standard'' Q-learning implementation requires a lookup table to store state-action values. Ipek et al. \cite{SelfOptMC2008} as well as Mukundan and Martinez \cite{MORSE2012} instead used $CMAC$ \cite{CMAC}, replacing a potentially extensive Q-learning table with multiple coarse-grain overlapping tables. This approach also included hashing, using hashed state attributes to index the CMAC tables. Taken together, these two methods balance generalization and overhead, although may introduce collisions/interference depending on the task. Further pipelining the hashing, CMAC table lookup, and calculation allows more possible action-values to be evaluated per cycle.


\textbf{Optimization}: Online ML applications with online training benefit from adaptivity to run-time workload characteristics. Despite these benefits, low model accuracy can negatively impact system performance, most notably at the start of execution or during periods of high variance in workload characteristics. Adaptations to control and learning can be considered to avoid these detrimental impacts. Some RL-based work \cite{Semantic2015} considered mitigating the impact of poor actions during exploration by introducing ``shadow'' operations. These operations are low confidence actions suggested by the model that are still used in model updates but not executed by the system. Consequently, the model gains feedback on the goodness of the action without negatively impacting the system. In a supervised learning based control task, Won et al. \cite{BootstrapCMP2014} trained an ANN online using control actions made by a PI controller, which exhibits far less start-up delay. Following training, control decisions are made using a hybrid combination based on error and consistency, allowing complementary control. In the simplest case, checking the performance of a default configuration, as in \cite{MCT2017}, provides a guarantee that the ML model will not perform worse than the default, but can perform better. 

In most works, ML models replace existing approaches (commonly a heuristic). Nevertheless, several recent works \cite{PPF2019, CNNHelperPred2019} have demonstrated significant advantages by combining both traditional (non-ML) and ML approaches. These improvements are derived from the orthogonal prediction/decision-making capabilities of the two approaches, thus enabling synergistic performance improvements. This method can also enable lower-cost ML application by focusing on particular shortcomings in traditional approaches. Both recent works \cite{PPF2019, CNNHelperPred2019} consider just branch prediction, thus significant opportunities exist to explore this potential co-design paradigm.



\subsection{Offline ML Applications}

\textbf{Model/Feature Selection}:
Offline ML applications generally exhibit substantial model/feature diversity since the model itself is not tied to a particular architecture. Model and feature selection therefore focuses more on maximizing model accuracy while minimizing overall learning/prediction time.
Design space exploration, in particular, can be approached using either iterative search methods for direct optimization or supervised learning methods to select optimal points based on the predicted optimality of a design. Several works \cite{3DNoC2015, Energy3DNoC2016, MOOSTAGE2018} used an iterative \textit{STAGE} \cite{STAGE} algorithm that optimizes local search for 3D NoC links by learning an evaluation function to predict local search results from a given starting point. Recent work has instead applied deep reinforcement learning \cite{DRL2019} to routerless NoC design. The proposed Monte Carlo tree search, along with actions suggested by a convolutional neural network, provide a highly efficient search process. Parallel threads are also utilized to scale design space exploration with increasing computational resources. System-level design space exploration has favored more standard supervised learning approaches \cite{MultiObjGPU2016, MLNoC2018, UncoreRPD2016}. Specific model choices vary, with linear \cite{MultiObjGPU2016, MLNoC2018} and non-linear \cite{UncoreRPD2016} regression models, as well as random forests and neural networks \cite{MLNoC2018} finding implementation. As in online ML applications, discussed in Section \ref{OnlineApp}, some tasks are naturally limited to supervised learning methods. Cross-architecture prediction is an exemplar \cite{CrossPlatPred2015, PhasePred2016, Stargazer2012, GPUPred2014, XAPP2015}. 





\textbf{Optimization}:
The usefulness of an ML model in offline ML applications is largely determined by overhead relative to traditional design approaches. Optimization therefore primarily focuses on improving data efficiency and overall model accuracy. 

Ensemble methods have been proposed in online ML applications \cite{MCT2017}, but primarily find application in offline ML applications as ensembles can be made arbitrarily large (relative to available computation resources). Several optimizations have been suggested to improve efficiency. Jooya et al. \cite{MultiObjGPU2016} trained many neural networks using slightly different configurations and generated an ensemble using a subset of the models that generalized well and were most insensitive to input noise. They further introduced outlier detection by filtering predictions whose performance and/or power predictions differ greatly from the closest configuration in training data. Ardalani et al. \cite{XAPP2015} instead kept all 100 models that they trained, noting that models may be very strong predictors in one application but weak predictors in another. They remedied this dilemma by selecting only the 60 closest individual predictions to the median prediction.

Sampling method optimization, while not unique to architecture tasks, are nevertheless important to consider in improving model accuracy. Sangaiah et al. \cite{UncoreRPD2016} considered potential systematic biases in their uncore performance prediction model. Specifically, they observed that uniform random sampling may not adequately capture performance relationships in a non-uniform configuration space (as in cache configurations using powers of two for sizing). They therefore used a low-discrepancy sampling technique, $SOBOL$ \cite{SOBOL}, to remove this systematic bias and prevent performance over-prediction for low-end configurations.

\subsection{Domain Knowledge \& Model Interpretation}
The powerful relationship learning capabilities offered by ML algorithms enable black-box implementation in many tasks (i.e., without consideration for task-specific characteristics), but may fail to capitalize on additional domain knowledge that could improve interpretability or overall model performance. Additionally, in some applications, domain knowledge can help identify aberrant behavior and, again, improve overall model usefulness. These themes are highlighted in several specific works, but can be generally applicable for ML applied to architecture.

One approach uses mechanistic-empirical models, synthesizing a domain knowledge based mechanistic framework with empirical ML based learning for specific parameters. These models simplify implementation compared to purely mechanistic models \cite{MechEmpModel2011}, can avoid incorrect assumptions made in purely mechanistic models \cite{SVR-NoC2013}, and can offer higher accuracy than purely empirical models by avoiding overfitting \cite{MechEmpModel2011}. Eyerman et al. \cite{MechEmpModel2011} also demonstrated how these models can be used to construct CPI stacks, allowing meaningful alternative design comparisons.

Deng et al. \cite{MCT2017}, in their work predicting optimal NVM write strategies, presented a case for tuning ML models based on task specific domain knowledge. Following initial analysis, they discovered how a single configuration parameter (wear quota) can result in higher complexity and sub-optimal prediction accuracy for IPC and system energy, even with quadratic regression and gradient boosting models. Excluding wear quota from the configuration space, then later applying it to the predicted optimal configuration, provided a 2-6\% improvement in prediction accuracy. Ardalani et al. \cite{XAPP2015} similarly examined inherent imperfections in their learning model for cross-platform performance prediction. Some predictions can be easy for learning models and hard for humans, representing an ideal scenario for ML application; the converse can also be true. In both cases, ML application is strengthened by considering task characteristics.




\section{Future Work}
This section synthesizes observations and analysis from Section \ref{Overview} and Section \ref{Analysis} to identify opportunities and detail the need for future work. These opportunities may come at the model level, exploiting improved implementation strategies and learning capabilities, or at the application level, addressing the need for generalized tools or exploring altogether new areas.

\subsection{Investigating Models \& Algorithms}
Existing works generally apply ML at a single time-scale or level of abstraction. These limitations motivate investigation into models and algorithms that capture the hierarchical nature of architecture, both in terms of system design and execution characteristics. 

\textbf{Perform Phase-level Prediction}: 
Application analysis using basic blocks \cite{BasicBlock2001} has long been a useful method for simulation, made possible by identifying unique and representative phases in program execution. Phase-level prediction offers analogous benefits for ML applied to architecture. 
A few recent works, in particular, have demonstrated promising results, with high accuracy for both performance prediction \cite{PhasePred2016} as well as energy and reliability (lifetime) \cite{MCT2017}. In general, most work \cite{LEO2015, MultiObjGPU2016, UncoreRPD2016} has not yet adopted phase-level prediction techniques (or does not explicitly mention their methodology). Specifically, future work could explore predictions for control and system reconfiguration based on phase-level behavior, rather than either static windows \cite{ManageCMP2008} or application-level behavior \cite{CALOREE2018, NumaPredict2016}.


\textbf{Exploit Nanosecond Scale}:
Coarse-grain ML, used in many DVFS control schemes, provides significant benefits over standard control-theory based schemes, yet fine grain control can provide even greater efficiency. Specifically, analysis by Bai et al. \cite{VoltRegRL2017} indicated very rapid changes in energy consumption, on the order of 1K instructions for some applications. Exploiting these brief intervals requires careful consideration for both the model and the algorithm. Future work may optimize existing algorithms such as experience sharing \cite{ExpShareRL2003} and hybrid/tandem control \cite{BootstrapCMP2014}, or consider approaches more suited for novel models (e.g., hierarchical models). These approaches could also enable additional nanosecond-scale co-optimization opportunities, such as dynamic LLC partitioning, to extract further efficiency gains.


\textbf{Apply Hierarchical \& Multi-agent Models}: 
Application execution in computer systems naturally follows a hierarchical structure in which, at the top level, tasks are allocated to cores, then cores are assigned dynamic power and resource budgets (e.g., LLC space), and finally, at the bottom level, data/control packets are sent between cores and memory. Consequently, a single machine learning model may struggle to learn appropriate design/control strategies. Furthermore, in the case of reinforcement learning models, it can be exceedingly difficult to accurately assign credits to specific low-level actions based on their impact on overall execution time, energy efficiency, etc. One promising approach in recent work is hierarchical models \cite{HierarchicalDRL2016}. Hierarchical reinforcement learning models enable goal-directed learning that is particularly beneficial in environments with sparse feedback (e.g., task allocation). Applying hierarchical learning to architecture could therefore enable more effective multi-level design and control. Multi-agent models are another promising area in machine learning research. These models tend to focus on problems in which reinforcement learning agents have only partial observability of their environment. Although partial-observability may not be a primary concern in individual computer systems, recent work \cite{MultiagentDRL2019} has applied this concept to internet packet routing and demonstrated convergence benefits via improved cooperation between individual agents.


\subsection{Enhancing Implementation Strategies} \label{sec:strats}
Increasingly complex models require effective strategies and techniques to reduce overhead and enable practical implementation. 
Model pruning and weight quantization, as discussed below, are two particularly effective techniques with proven benefits in accelerators, while many other promising approaches are also being explored \cite{AccelSurvey2017}. 

\textbf{Explore Model Pruning}: 
Model complexity can be a limiting factor in online ML applications. A standard Q-learning approach requires a potentially extensive table to store action-values. Neural network based approaches for both RL (in Deep Q-Networks) and supervised learning require network weight storage and additional processing capabilities. Neural networks, in particular, are therefore generally constrained to a few layers in existing work, with many using just one hidden layer \cite{ANNNoCLinks2012, BootstrapCMP2014, ManageCMP2008, MITHRA2016} and some using one or two hidden layers \cite{NPU2012, NGPU2015}. 

Recent research on neural networks has demonstrated promising methods to reduce model complexity through pruning \cite{EfficientNN2015, SparseTrain2018}. The general intuition is that many connections are unnecessary and can therefore be pruned. Iteratively pruning a high-complexity network, then re-training from scratch on the sparse architecture achieves good results, with some work demonstrating very high sparsity ($>$90\%) and little accuracy penalty \cite{SparseTrain2018}.

Pruning applied to neural networks, either in deep Q-learning or supervised learning regression/classification, offers a method to train complex models for high accuracy, then prune for feasible implementation. Deep Q-learning application has, thus far, been limited to two works \cite{LEADRL2019, DRLArbitration2018}, one of which is currently impractical to implement \cite{DRLArbitration2018}. Future work may instead consider pruned deep Q-networks as a useful alternative to standard Q-learning approaches. Pruning also provides a substantial opportunity for future work on performance prediction (as in DVFS control) and function approximation (as in ML-enabled approximate computing). System-level approximation (discussed in Section \ref{sec:applications}) may particularly benefit from pruning high complexity models.

\textbf{Explore Quantization}: 
Existing work primarily applies quantization to state values in Q-learning to enable practical Q-table implementation. Similarly, neural networks benefit from potential reduction in execution time, power, and area by reducing multiply-accumulator precision. Recent works, however, suggest a new spectrum of opportunities for alternative hardware implementations based on reduced precision models. 

Binary neural networks, for example, quantize weights to be either +1 or -1, enabling computation based on bit-wise operations rather than arithmetic operations \cite{BinaryNN2016}. An additional approach considered quantizing neural network weights into finite (but non-binary) subsets in order to replace multiply operations with lookup-table accesses \cite{LookNN2017}, allowing higher precision and lower execution time, albeit with potentially higher area cost. Future work on ML application can exploit similar hardware implementations while exploring optimal quantization levels for various tasks and control schemes.




\subsection{Developing Generalized Tools}
Existing machine learning tools (e.g., scikit-learn \cite{scikit-learn}) have proven useful for relatively simple ML applications. Nevertheless, complex design and simulation tasks require more sophisticated tools to enable rapid task-specific optimizations using general-purpose frameworks.

\textbf{Enable Broad Application \& Optimization}: 
Purpose-built architectural tools, similar to heuristic design strategies, can be useful in enabling design, exploration, and simulation that satisfies a common use case. These approaches may still be limited in their application to a specific problem, optimization criteria, system configuration, etc. Given the fast-paced nature of architectural research (and machine learning research), there is a need to develop more generalized tools and easily modifiable frameworks to address broader applications and optimization options. 


ML-based design tools are especially promising, with recent works demonstrating successful application to immense design spaces (e.g., exceeding $10^{12}$ in \cite{DRL2019}). Opportunities for new design tools are not, however, limited to specific architectural components. Chip layout is a notable example in which even simple clustering algorithms can dramatically outperform existing heuristic approaches \cite{FFCluster2016}. Future work can also continue to develop more broadly applicable tools for performance and power prediction. In particular, recent work on cross-platform performance prediction \cite{SXAPP2019} suggests the possibility for high prediction accuracy with purely static features, thus representing another potential area for additional research.

\textbf{Enable Widespread Usage}: 
Generalized tools enable additional benefit by facilitating rapid design and evaluation. Using a machine learning approach, one might simply modify training data (in a supervised learning setting) or action/reward representation (in a reinforcement learning setting) rather than exploring models, data representation strategies, search approaches, etc., possibly without \textit{a priori} machine learning experience. For example, recent work \cite{DRL2019} envisioned reuse of a deep reinforcement learning framework for diverse NoC-related design tasks involving interposers, chiplets, and accelerators. While the framework might not be compatible with all work, especially in novel areas, it may provide a better foundation for machine learning application to architectures, especially for individuals with limited machine learning background.





\subsection{Embracing Novel Applications} \label{sec:applications}
Opportunities abound for future work to apply ML to both existing and emerging architectures, replace heuristic approaches to enable long-term scaling, and advance capabilities for automated design.

\textbf{Explore Emerging Technologies}:
Several proposals \cite{SSDCacheML2018, RLTailLatency2018, MCT2017, DiskFailPred2018} establish how ML can be used to optimize both standard (energy, performance) and non-standard (lifetime, tail-latency) criteria. These non-standard criteria are shown to be particularly problematic in emerging technologies as these technologies cannot easily find widespread implementation without some reliability guarantees. 
Applying ML to optimize both standard and non-standard criteria therefore provides a method for future work to intelligently balance control strategies dynamically, rather than relying upon a heuristic approach.

\textbf{Explore Emerging Architectures}: 
ML application to emerging architectures presents a similar benefit by enabling rapid development, even with limited best-practice knowledge, which may take time to develop. Work in long-standing design areas, such as task allocation and branch prediction, may incorporate best-practice domain knowledge to guide approaches, whether applying ML or some other traditional method. Best practices for emerging architectures may not be immediately obvious. For example, ML application to 2D photonic NoCs \cite{Photonic2018}, 2.5D processing-in-memory designs \cite{GPUPIM2016}, and 3D NoCs \cite{3DNoC2015, Energy3DNoC2016, MOOSTAGE2018} have all shown strong performance over existing approaches. Future work can explore ML application to novel concerns such as connectivity and reconfigurability in interposers and domain-specific accelerators.

\textbf{Expand System-Level Approximate Computing}: 
As discussed in Section \ref{MLapprox}, ML applications for approximate computing have been mostly limited to function approximation. 
However, there are many other facets of approximate computing that have already been implemented in non-ML works, which can be reap additional benefits by utilizing ML.
For example, APPROX-NoC \cite{APPROX-NoC2017} reduces network traffic using approximated and encoded data. Another work explored a multi-faceted approximation scheme for a smart camera system \cite{SmartCamApprox20107} using approximate DRAM (lower refresh rate), approximate algorithms (loop skipping) and approximate data (lower sensor resolution). Existing compiler-based work \cite{ACCEPT2015} for system-wide approximation enhances prior capabilities to determine approximable code, but relies upon heuristic searches with representative inputs. Consequently, this method does not provide statistical guarantees, such as those in MITHRA \cite{MITHRA2016}. Future work may explore searches based on deep reinforcement learning (or perhaps hierarchical reinforcement learning) to incorporate existing approximation techniques into a scalable framework for high-dimensional approximation and co-optimization.

\textbf{Implement System-Wide, Component-Level Optimization}:
Recent work has begun to explore broader ML-based design and optimization strategies. MLNoC \cite{MLNoC2018} explores a wide SoC feature space for NoC design optimization. Core and uncore DVFS are combined in Machine Learned Machines \cite{MLM2016}, along with LLC dynamic cache partitioning to explore co-optimization potential at run-time. Related DNN accelerator research \cite{AccelBayesian2017} proposed co-optimization of hardware-based (e.g., bitwidth) and neural network parameters (e.g., L2 regularization). These works motivate consideration for system-wide, component-level ML application.

Existing system-level optimization schemes (e.g., \cite{MLScheduling2018, HeterogeneousML2017, NumaPredict2016}) consider configuration opportunities at just a single and very high level of abstraction (e.g., task allocation or big.LITTLE core configurations). Although these works may include low-level features such as NoC utilization and DRAM bandwidth in their ML models, they do not account for the impact of component-level optimization techniques such as NoC packet routing, cache prefetching, etc. We instead envision an ML-based system-wide, component-level framework for run-time optimization. In this framework, control decisions would involve a larger hierarchy of both component-level (or lower) features and control options as well as higher-level decisions, allowing a more comprehensive and precise perspective for run-time optimization.


\textbf{Advance Automated Design}: 
While fully automated design might be the ultimate objective, increasingly automated design is nevertheless an important milestone for future work. Specifically, as more tasks are automated, there is greater potential to enable a positive-feedback loop between machine learning and architecture, providing immense practical benefits for both fields. There are, of course, a number of intervening challenges that must be solved, each of which represents a substantial area for future work.

One challenge involves modeling the hierarchical structure of architectural components. This model would likely benefit from integrating pertinent characteristics across the system stack, from process technology to full-system behavior, thus generating a highly accurate representation for real-world systems. Another research direction could explore methods for machine learning models to identify potential design aspects for improvement. Ideally, this model could explore not just reconfiguration of pre-existing options (as in \cite{CrossLayerReliability2016}), but also generate novel configuration options. Integrating these and potentially other capabilities may provide a framework to advance automated design.


\section{Conclusion}
Machine learning has rapidly become a powerful tool in architecture, with established applicability to design, optimization, simulation, and more. Notably, ML has already been successfully applied to many components, including the core, cache, NoC, and memory, with performance often surpassing prior state-of-the-art analytical, heuristic, and human-expert strategies. Widespread application is further facilitated by diverse training methods and learning models, allowing effective trade-offs between performance and overhead based on task requirements. These advancements are likely just the beginning of a revolutionary shift in architecture. 

Optimization opportunities at the model level involving pruning and quantization offer broad benefits by enabling more practical implementation. Similarly, opportunities abound to extend existing work using ever-more-powerful ML models, enabling finer granularity, system-wide implementation. Finally, ML may be applied to entirely new aspects of architecture, learning hierarchical or abstract representations to characterize full system behavior based on both high and low level details. These extensive opportunities, along with yet to be envisioned possibilities, may eventually close the loop on highly (or even fully) automated architectural design.





\bibliographystyle{ieeetr}
\bibliography{ref}


\end{document}